\documentclass[a4paper]{article} 
\usepackage[hyphens]{url}  
\usepackage{graphicx} 
\usepackage{graphicx}  

\usepackage{xcolor}
\usepackage{tikz}
\usepackage[ruled,linesnumbered]{algorithm2e}
\usepackage{amsmath}
\usepackage{amsfonts}
\usepackage{subcaption}
\usepackage{natbib}

\usetikzlibrary{positioning,shapes,calc,through,intersections,fit,matrix}

\tikzset{
  flowchart/.style={
      process/.style={rounded rectangle, draw,align=center},
      decision/.style={diamond, aspect=1.4, draw,align=center, inner
        sep=0.1em},
      transition/.style={draw,->},
      dataflow/.style={->,dashed},
    },
    mnemonic/.style={
      label={[fill=blue!30]100:#1}
    }
}

\begin{document}
%

\title{Double-oracle sampling method for Stackelberg Equilibrium approximation in general-sum extensive-form games}
\author{Jan Karwowski\\ Warsaw Univeristy of Technology\\
  Koszykowa 75, 00-662 Warsaw, Poland\\
  jan.karwowski@mini.pw.edu.pl
  \and Jacek Ma\'ndziuk\\
  Warsaw Univeristy of Technology\\
  Koszykowa 75, 00-662 Warsaw, Poland\\
  j.mandziuk@mini.pw.edu.pl
}
\date{}
\maketitle


\begin{abstract}
The paper presents a new method for approximating Strong Stackelberg Equilibrium in general-sum sequential games with imperfect information and perfect recall. The proposed approach is generic as it does not rely on any specific properties of a particular game model. The method is based on iterative interleaving of the two following phases: (1) guided Monte Carlo Tree Search sampling of the Follower's strategy space and (2) building the Leader's behavior strategy tree for which the sampled Follower's strategy is an optimal response.
The above solution scheme is evaluated with respect to expected Leader's utility and time requirements on three sets of interception games with variable characteristics, played on graphs. A comparison with three state-of-the-art MILP/LP-based methods shows that in vast majority of test cases proposed simulation-based approach leads to optimal Leader's strategies, while excelling the competitive methods in terms of better time scalability and lower memory requirements.
\end{abstract}


\section{Introduction}%
\label{sec:introduction}
Stackelberg Equilibrium (SE)~\cite{Leitmann_1978} defines equilibrium profile for two-player asymmetric games. One player -- the Leader -- commits to a certain strategy and the other player -- the Follower -- defines his/her strategy being aware of the Leader's commitment. The notion of SE, which originated in the field of economy, gained momentum in recent decade thanks to intensive research on Security Games~\cite{sinha2018securitygames} which often use Stackelberg Game (SG) to model interactions between a defender (playing the role of a Leader) and an attacker (being a Follower). We consider the Strong Stackelberg Equilibrium (SSE)~\cite{Leitmann_1978} in which (additionally to SE) the Follower breaks ties in favor of the Leader when calculating the optimal response.

Majority of contemporary SG research is focused on developing effective methods for specific games,
e.g.~\cite{brazdil2018patrolling-internet,schlenker2016get,basilico2012,xinrun2018catching,johnson2012patrol} and there are just a few works related to finding SE in the case of general SG models.
%
%
\subsection{Contribution}%
\label{sec:contribution}
The main contribution of this paper is a method for approximating SE in a broad and general genre of sequential general-sum imperfect-information games,
inspired by a double-oracle approach~\cite{bosansky2014doubleoracle,jain2011double}.

Despite being rooted in the double-oracle framework, the proposed method presents an entirely different operational principle than those of~\cite{bosansky2014doubleoracle,jain2011double} as it relies on iterative Monte Carlo Tree Search (MCTS)~\cite{mctsSurvey} sampling of the Follower's strategy alternated with an adjustment of the Leader's behavior strategy represented in the form of a tree.

Proposed method is experimentally proven to yield close-to-optimal defender's strategies while scaling better in time and memory usage than competitive MILP (Mixed Integer Linear Program) based approaches.

To the best of our knowledge there is only one other approach that utilizes MCTS method to solve general-sum extensive-form SGs~\cite{KarwowskiMandziuk2015,KarwowskiMandziuk2016,KarwowskiMandziuk_2019} which, however, adopts a different protocol and relies on iterative adjustment of the Leader's strategy by means of direct MCTS sampling against gradually changing Follower's strategy.
This method, though, could not be applied to solve games with complex information set (IS) structures, e.g. Search Games~\cite{bosansky2015} considered in this paper.

\subsection{Related Work}%
\label{sec:related-work}
In the literature, the problem of finding SE is usually considered in the context of some particular game model and therefore majority of proposed approaches are model-specific and cannot be straightforwardly applied to other kinds of SGs. On a general note, existing solution methods
usually adapt and tune one of the following
well-established techniques:
\emph{column and constraint generation} -- e.g.~\cite{xinrun2018catching,jain2010security}; \emph{marginal and compact strategies} -- exploiting a particular structure of a game and its payoffs, e.g.~\cite{kiekintveld2009computing,schlenker2016get}; or \emph{game abstraction} -- e.g.~\cite{xinrun2018catching,basak2016abstraction}.
Utilization of these techniques requires tailoring a solution method
to characteristic game properties, what leads to a highly efficient,
though game-dedicated algorithm.

An efficient exact approach to generic sequential general-sum SGs was proposed in~\cite{bosansky2015} where the authors considered a sequence-form representation of a sequential game to improve scalability of a corresponding MILP. Another powerful general approach, introduced by~\citeauthor{cermak2016using} (\citeyear{cermak2016using}) starts off with finding Stackelberg Extensive Form Correlated Equilibrium of a game using MILP and then restricts it iteratively until the obtained strategy profile
corresponds to SE. 
Yet another general approach to extensive-form games~\cite{CBK2018} starts from a smaller (restricted) game and gradually expands the game tree to compute the SSE.
These three state-of-the-art generic methods are used as reference points in experimental evaluation of the approximate approach proposed in this paper.

Our method (referred to as \emph{O2UCT} -- double-oracle UCT sampling) relies on a guided sampling of the Follower's strategy space and finding a feasible Leader's strategy using double-oracle method and does not involve solving Linear Program (LP) of any kind. Application of \emph{O2UCT} leads to scalability performance boost similar to that of using column and constraint generation method in LP/MILP, albeit \emph{with no direct reference to any specific game model properties, in the solution method}.


\section{Imperfect Information Stackelberg Games}%
\label{sec:stackelberg-games}
%
A sequential non-zero-sum game with imperfect information can be defined using an \emph{extensive form}.
An Extensive Form (EF) game is an $8$-tuple $G=(\mathcal{N},
\mathcal{S}, \mathcal{Z}, \rho, \mathcal{A}, u, \mathcal{T},
\mathcal{I})$, where $\mathcal{N}=\{L,F\}$ is a set of players (the
Leader and the Follower in the case of a two player SG). $\mathcal{S}$
and $\mathcal{Z}$ are sets of non-terminal and terminal game states,
resp., $\rho: \mathcal{S} \rightarrow \mathcal{N}$ is a function defining
which player acts in a given state. $\mathcal{A}=\bigcup_{s\in S}
{A_s}$ is a family of sets $A_s$, where $A_s$ is a set of all actions
available to an active player in state $s$. $u :\{\mathcal{Z} \times
\mathcal{N}\} \rightarrow [0,1]$ is a utility function which provides
utilities for all players in terminal states. $\mathcal{T}$ is a set
of transition functions $T_s: A_s \rightarrow \mathcal{S} \cup
\mathcal{Z}$ such that for every $s\in \mathcal{S}$, $T_s(a)$ is a state resulting from playing action $a$ in state $s$. $\mathcal{I}$ is a family of information sets $I_{k}\subseteq \mathcal{S}$ which satisfies the standard definition~\cite{kuhn1950extensive}.

Moreover, all considered games have \emph{perfect recall} property, i.e. an active player is fully aware of his/her past actions and ISs he/she encountered before reaching the current state.

Let's denote by $A_{I_k}$ a set of actions available in a given information set $I_k$ and by $\mathcal{I}^{n}$ a family of information sets in which player $n$ is an active player. A \emph{pure strategy} of player $n$ (denoted by $\pi_n$) is an assignment of one of the allowed actions per each IS in $\mathcal{I}^{n}$. A \emph{mixed strategy} $\delta_n$ is a probability distribution over all possible pure strategies $\pi_n$ of player $n$. In EF games a \emph{behavior strategy} is additionally defined, as a function that assigns a probability distribution over all available actions to each IS. In \emph{perfect recall} games mixed and behavior strategies are pairwise equivalent~\cite{kuhn1950extensive} and therefore in the remainder of the paper we will denote behavior strategies of player $n$ by $\delta_n$ and treat them equivalently to mixed strategies.

We will use the notation $\mathbb{E}U^n_{\delta_F,\delta_L}$ to denote the expected utility of player $n$ ($L$ -- Leader, $F$ -- Follower) when the Leader and the Follower play strategies $\delta_L$ and $\delta_F$, respectively. An index referring to the Leader's strategy will be omitted in the contexts in which it does not lead to misunderstandings.




\section{Double-oracle sampling method (\emph{O2UCT}) for SE approximation}%
\label{sec:mixed-method}

\begin{figure}
  \centering
  \begin{tikzpicture}[
    flowchart,
    impl/.style={dashed,draw,align=center}
  ]
  \node(sample)[process,inner sep=2pt,label={[label
    distance=-.9em]160:Step~1}]{Sample Follower's strategy\\ in a
    guided manner};
  \node(sampleimpl)[impl,below=of sample,yshift=2.85em]{Apply UCT selection,\\ expansion \& simulation\\
    to AFG};

\node[above=of sample, xshift=4em,yshift=-1.5em]{Repeat for a pre-defined number of iterations:};

  \node(train)[process, yshift=1.5em,below = of sampleimpl,
  thick,fill=gray!20,label={[label
    distance=-.9em]150:Step~2}]{Calculate Leaders's\\ strategy};
  \node(trainimpl)[impl, xshift=5em, below=of train,yshift=2.85em]{Consider moves
    on the path from the root to\\ the leaf in the AFG to be a pure
    Follower's strategy\\ and apply procedure from Section~\ref{sec:leaders-method}};
  \node(collect)[process, right=of sample,xshift=-2.4em,inner sep=2pt]{Collect game payoffs\\ and
    use them to guide\\ subsequent sampling};
  \node(collectimpl)[impl, yshift=2.85em,below=of collect,
  label={[label distance=.2em]300:Step~3}]{Backpropagate the
      Leader's\\ payoff in the UCT tree.};

  \draw[transition] (sampleimpl)--(train);
  \draw[transition] (train)-|(collectimpl) node[pos=0.6, auto]{calculate payoffs};
  \draw[transition] (collect)--(sample);
\end{tikzpicture}
  \caption{An outline of the \emph{O2UCT} method. Oval frames present the method's general idea, while rectangular dashed frames summarize particular realization of each step proposed in this work. Implementation of Step~2, which is the most challenging part of \emph{O2UCT}, is described in more detail in Fig.~\ref{fig:mixed-method-overview}.
  }
  \label{fig:method-outline}
\end{figure}
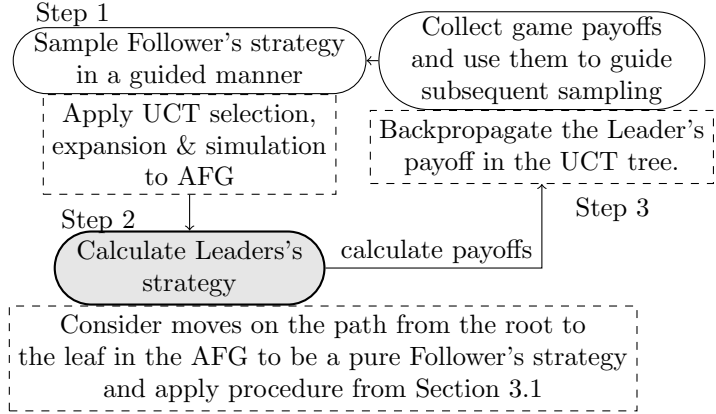

In order to find SE in a perfect recall imperfect-information deterministic multi-act general-sum game the following iterative procedure, depicted in Fig.~\ref{fig:method-outline}, is applied. In each iteration, in the first step the Follower's strategy is sampled with a method capable of using the results from previous iterations to guide subsequent sampling. Next, a method for finding the Leader's strategy, {\bf for which the just-sampled Follower's strategy is the optimal response}, is applied. In the third step utility values corresponding to obtained strategy profile are collected to adjust the guided sampling procedure (Step~1) in the next iteration.

A distinctive feature of the proposed method is the lack of exhaustive
search of the Follower's strategy space, which is replaced by an
iterative guided space sampling procedure.

\begin{figure}
  \centering
  \begin{tikzpicture}[
    isf/.style={rectangle,draw},
    isl/.style={circle,draw},
    level 1/.style={sibling distance=5.5em},
    level 2/.style={sibling distance=2.7em, level distance=3em},
    level 3/.style={sibling distance=1.5em, level distance=5em},
    afgnode/.style={draw,diamond, inner sep=0.1em, minimum width=3em,
      minimum height=3em},
    nn/.style={inner sep=0},
    move/.style={pos=.5, left},
    terminal/.style={regular polygon, regular polygon sides=8, draw,
      inner sep=.17em,fill=red!5}
    ]
    \node[isf] {$I_1$}
    child {
      node[isl] {} child {
        node [isf] {$I_2$} child {
          node [terminal] {$z$}
        } child {
          node[nn] {$\vdots$}
        }
      } child {
        node [isf] {$I_3$} child {
          node[terminal] {$z$} edge from parent node[move, pos=0.4] {$m_3$}
        } child {
          node[terminal] {$z$} edge from parent node[move,right, pos=0.4] {$m_4$}
        }
      } edge from parent node[move] {$m_1$}
    } child {
      node[isl] {} child {
        node(f21) [isf] {$I_4$} child {
          node[nn] {$\vdots$}
        } child {
          node[terminal] {$z$}
        }
      } child {
        node(f22) [isf, xshift=0.3em] {$I_4$} child {
          node[terminal] {$z$}
        } child {
          node[nn] {$\vdots$}
        }
      } edge from parent node[move] {$m_2$}
    };

    \draw[ultra thick, gray, double,->] ((4em, -3em) -- (9em, -3em);

    \begin{scope}[xshift=14em,
      level 1/.style={sibling distance=6em},
      level 2/.style={sibling distance=4em, level distance=5em},
      ]
      \node(qq1)[afgnode] {$I_1'$} child {
        node(qq2)[afgnode] {$I_2'$} child {
          node(qq3)[afgnode] {$I_3'$} child {
            node[terminal] {$\,$}  edge from parent node[move] {$m_3'$}
          } child {
            node[terminal] {$\,$}  edge from parent node[move,right] {$m_4'$}
          }
        } child {
          node[xshift=4em] {$\cdots$} child[level distance=3em, sibling distance=3.5em] {
            node[afgnode] {$I_3''$} child [sibling distance=1.5em,solid,thin] {
              node [terminal] {$\,$} edge from parent node[move] {$m_3''$}
            } child [sibling distance=1.5em,solid,thin] {
              node [terminal] {$\,$} edge from parent node[move,right] {$m_4''$}
            }            edge from parent [dotted, thick]
          } child[level distance=3em, sibling distance=3.5em] {
            node[afgnode] {$I_3'''$} child[sibling distance=1.5em, solid,thin] {
              node [terminal] {$\,$}
            } child[sibling distance=1.5em,solid,thin] {
              node [terminal] {$\,$}  edge from parent node[move,right] {$m_4'''$}
            }             edge from parent [dotted, thick]
          } child[level distance=3em, sibling distance=3.5em] {
            node[xshift=-1em] {$\cdots$} edge from parent [dotted, thick]
          }
        } edge from parent node[move] {$m_1'$}
      } child {
        node[afgnode, xshift=1em, yshift=2em] {$I_4'$} child[sibling distance=2em,level distance=3em] {
          node[nn] {$\vdots$}
        } child [sibling distance=2em, level distance=3em] {
          node[nn] {$\vdots$}
        } edge from parent node[move,pos=0.4,auto] {$m_2'$}
      };
    \end{scope}

    \draw[dashed] (f21) -- (f22);

    \begin{scope}[
      yshift=-18em,
      queue/.style={draw=gray},
      dashed,
      queuelink/.style={dotted,draw=black!50!gray}
      ]

      \node(q1)[queue,xshift=-1em,yshift=-1em] {
        $I_1$
      };

      \node(q2)[queue,right= of q1, xshift=6em] {
        $I_2,I_3$
      };

      \node(q3)[queue,right= of q2, xshift=2em] {
        $I_3$
      };

      \draw[queuelink] (q1) .. controls ++(9em,3em) and ++(-7em, -1em) .. (qq1);
      \draw[queuelink] (q2) .. controls ++(-3em, 2em) and ++(-7em, -7em)
      ..(qq2);
      \draw[queuelink] (q3) .. controls ++(-8em, 2em) and ++(4em, -4em)
      ..(qq3);

      \draw[->] (q1) -- (q2) node [pos=0.5,auto] {$-I_1$, $+\{I_2,I_3\}$};
      \draw[->] (q2) -- (q3) node [pos=0.5,auto] {$-I_2$, $+\emptyset$};

    \end{scope}
  \end{tikzpicture}
  \caption{An example of AFG construction. Original two-player EF game
    (on the left) and the resulting one-player AFG (on the
    right). Rectangles in the bottom part of the figure present a state of a queue in states $I_1'$, $I_2'$ and $I_3'$, respectively assuming that a path in AFG which is currently considered by the algorithm is the leftmost one. Signs $+$ and $-$ represent \emph{push} and \emph{pop} operations in the queue, resp.
}%
  \label{fig:afg}
\end{figure}
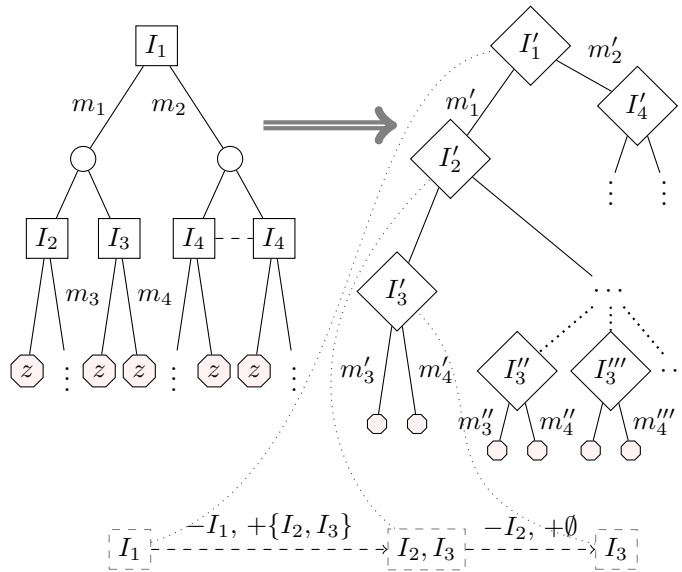

In order to perform this sampling an Auxiliary Follower's Game (AFG)
is formulated, which is a \emph{one-player} game that yields the
Follower's \emph{restricted pure strategy} (also called
\emph{restricted pure realization plan} in
EF games) when reaching a terminal state of AFG. AFG is constructed
based on the original EF game in the following way:
  \begin{itemize}
  \item The current AFG state is represented in the form of a queue of the Follower's ISs (from the original ES game). Initially the queue contains ISs observable by the Follower before their first move.
  \item Each game round consists in taking the first IS from the queue, playing one of the moves available in that IS and placing in the queue all ISs that may results from the current IS after playing the selected move (i.e. all ISs for which there exists a Leader's strategy directly leading to them).
  \item The game is played until the queue is empty.
  \end{itemize}
  The moves played on the path from the root of AFG to its leaf
  define a pure Follower's strategy in the original EF
  game. An example transformation of two-player EF game to one-player AFG is presented in Fig.~\ref{fig:afg}.

%

In principle, there are no formal requirements for a sampling
heuristic method to be used, except for the ability to transfer
knowledge related to the sampled space to subsequent iterations. In
the experiments an MCTS~\cite{mctsSurvey} variant called Upper
Confidence Bound applied to Trees (UCT) proposed
by~\citeauthor{UCT}~\cite{UCT} was applied to guide the Follower's
strategy space
sampling process by means of finding the optimal strategy in AFG,
as described in the next section.

In short, each MCTS/UCT iteration (playout) is composed of $4$ main phases: \emph{selection}, \emph{expansion}, \emph{simulation}, and
\emph{backpropagation} (please consult~\cite{UCT} or~\cite{mctsSurvey} for a detailed description). In our method, selection, expansion and
simulation correspond to the first step of an \emph{O2UCT} iteration (guided sampling) and backpropagation phase is implemented in its third step
(collection of payoffs). The second step in Fig.~\ref{fig:method-outline} refers to reaching the leaf node (a final state of the AFG) and obtaining the Leader's payoff. This payoff is equal to the expected payoff of playing the Leader's strategy and is calculated using a method presented in Section~\ref{sec:leaders-method}.

UCT is a powerful and versatile metaheuristic which has proven successful in a wide variety of optimization problems, including General Game Playing~\cite{swiechowski2016ggp}, playing classical board games~\cite{alphago-zero}, proactive planning under uncertainty~\cite{WaledzikMandziuk2018} or combinatorial optimization~\cite{sabharwal2012guiding}.

The most challenging part of \emph{O2UCT}
is an algorithm (described below) which for the current Follower's strategy $\pi_f$ finds the respective Leader's strategy $\delta_l$ for which  $\pi_f$ is an optimal response. The final outcome of \emph{O2UCT} is a pair of strategies $(\delta_l, \pi_f)$ providing the highest Leader's payoff found across all iterations.

\subsection{A method of finding the Leader's strategy}%
\label{sec:leaders-method}

The method utilizes a tree structure representation of the Leader's behavior strategy which has the following properties.
\begin{itemize}
\item Each node is labeled with the Leader's IS and contains a vector of probabilities of actions available in this IS.
\item Root node represents the initial Leader's IS.
\item Edges going out of any node are labeled with pairs $(a,I')$, where $a$ is an action and $I'$ is an IS reachable by playing $a$ in a given node.
\end{itemize}
Please note that several ISs may be reachable after
playing the same move, depending on an opponent's response. Such a
situation is depicted in Fig.~\ref{fig:mixed-tree} presenting an
example tree in which playing move $m6$ may lead to either one of the
two ISs ($s9$ or $sa$) depending on a move played by the
opponent. Initially, the tree does not contain all ISs,
but only those reachable by an initial strategy profile. Subsequent nodes are added gradually, as explained below.

The algorithm for finding the Leader's strategy is inspired by a double-oracle approach~\cite{bosansky2014doubleoracle,jain2011double} and consists of alternating the following two phases: (1) an improvement of the Leader's strategy against a fixed Follower and (2) finding the optimal Follower's response against the current Leader's strategy -- based on the Follower's oracle.
%
For a sampled Follower's strategy (Step~1 in Fig.~\ref{fig:method-outline}) a corresponding Leader's strategy (Step~2 in Fig.~\ref{fig:method-outline}) must satisfy the following conditions:
\begin{description}
\item[(*)] the optimal Follower's response to that strategy is the
  same as the sampled Follower's strategy,
\item[(**)] among all Leader's strategies that satisfy the above constraint it is the one that optimizes the Leader's payoff.
\end{description}
Any Leader's strategy satisfying (*) will be called a \emph{feasible strategy} (a set of \emph{feasible strategies} is also called a \emph{best response region} in~\cite{vonstengel2004leadership}).

Let us denote the sampled Follower's strategy by $\pi_F^r$ ($r$
stands for the \emph{requested} Follower's strategy). An overview of the method of finding the Leader's strategy that fulfills constraints (*)--(**) is presented in Fig.~\ref{fig:mixed-method-overview} and consists of the following steps:
\begin{enumerate}
\item Initialize the Leader's strategy.
\item Seek the Follower's strategy yielding better Follower's payoff against the current Leader's strategy using the algorithm described in Section~\ref{sec:oracle}. If such strategy exists call
  it $\pi^b_F$ ($b$ stands for \emph{better} (in terms of payoff) Follower's strategy).\label{item:bas}
\item If $\pi^b_F$ was found, then perform strategy \emph{feasibility pass} (see below) and go to~\ref{item:bas}, otherwise continue.
\item If stopping condition is not met, perform the Leader's strategy adjustment that increases the Leader's payoff (\emph{positive pass} - see below) and go to~\ref{item:bas}, otherwise continue.\label{item:di}
\item Return the best Leader's strategy among all \emph{feasible strategies} found in step (3).
\end{enumerate}

%
\begin{figure}[t]
  \centering

    \begin{tikzpicture}[
  msnode/.style={shape=circle,draw,minimum width=2.5em},
  fork/.style={draw,shape=circle,fill, inner sep=0, minimum width=.15em},
  edge/.style={draw},
  node distance=2.0em,
  ]
  \node[msnode](root){s1};

  \node[fork, label=above:{$m1,0.8$}, below=of root, xshift=-4em](m1){};
  \node[msnode,below=of m1, xshift=1em](s2){s2};
  \node[msnode,below=of m1, xshift=-3em](s3){s3};

  \draw[edge] (root)--(m1);
  \draw[edge] (m1) .. controls ++(-.5em,-.5em) .. (s3);
  \draw[edge,draw=green] (m1)  .. controls ++(-.5em,-.5em) .. (s2);

  \node[fork,label=above:{$m3,1.0$}, below= of s3, xshift=-2em](m3){};
  \node[msnode, below= of m3, xshift=-2em](s5){s5};
  \draw[edge] (s3)--(m3)--(s5);

  \node[fork, label=above:{$m4,0.5$}, below=of s2, xshift=-2em](m4){};
  \node[msnode, below=of m4, xshift=-2em](s6){s6};
  \node[msnode, below=of m4, xshift=2em](s7){s7};
  \draw[edge] (s2)--(m4);
  \draw[edge] (m4)--(s6);
  \draw[edge,draw=green] (m4) .. controls ++(-.25em, -.5em) .. (s7);

  \node[fork, label=above:{$m5,0.5$}, below=of s2, xshift=2em](m5){};
  \node[msnode, below=of m5, xshift=1em](s8){s8};
  \draw[edge] (s2)--(m5)--(s8);

  \node[fork, label=above:{$m2,0.2$}, below=of root](m2){};
  \node[msnode,below=of m2](s4){s4};
  \draw[edge] (root)--(m2);
  \draw[edge] (m2)--(s4);

  \node[fork, label=above:{$m6,0.0$}, below=of root, xshift=6em](m6){};
  \node[msnode, below=of m6, xshift=-3em](s9){s9};
  \node[msnode, below=of m6, xshift=3em](sa){sa};

  \draw[edge] (root) -- (m6);
  \draw[edge] (m6) .. controls ++(.5em, -.5em) .. (s9);
  \draw[edge,color=green] (m6) .. controls ++(.5em, -.5em) .. (sa);

  \node[fork,label=above:{$m7,1.0$}, below=of s9,xshift=-0em](m7){};
  \node[msnode,below=of m7](sb){sb};
  \draw[edge] (s9)--(m7)--(sb);

  \node[fork,label=above:{$m8,0.3$}, below=of sa,xshift=-2em](m8){};
  \node[fork,label=above:{$m9,0.4$}, below=of sa,xshift=+2em](m9){};
  \node[fork,label=above:{$ma,0.3$}, below=of sa,xshift=+6em](ma){};
  \node[msnode,below=of m8](sc){sc};
  \node[msnode,below=of m9](sd){sd};
  \node[msnode,below=of ma](se){se};
  \draw[edge] (sa) .. controls ++(-2em,-1em) .. (m8) -- (sc);
  \draw[edge] (sa) .. controls ++(2em,-2em) .. (m9) -- (sd);
  \draw[edge] (sa) .. controls ++(6em,-2em) .. (ma) -- (se);

\end{tikzpicture}
    
  \caption{An example of the Leader's behavior strategy tree. Nodes represent the Leader's ISs. Edges are labeled with pairs (move, probability). Playing some moves (e.g. $m6$) may lead to various ISs ($s9$ or $sa$, resp.), depending on the Follower's action.}%
  \label{fig:mixed-tree}
\end{figure}
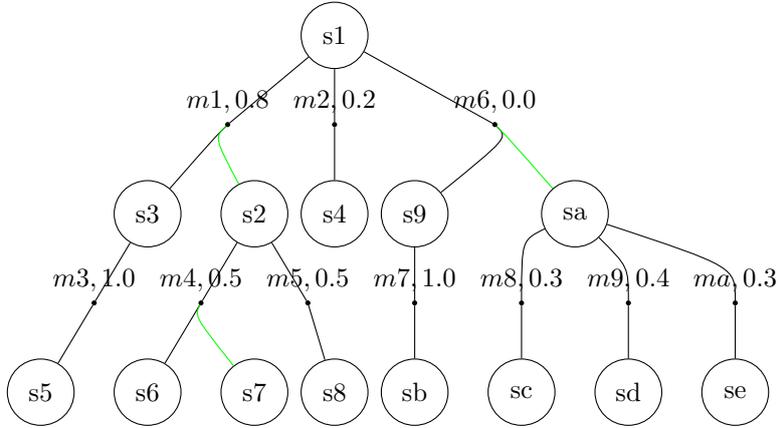
\begin{figure}[t]
  \centering
  \begin{tikzpicture}[
  flowchart,
  dsmutate/.style={fill=red!30}
  ]

  \node(init)[process] {Initialize Leader's\\ mixed strategy};
  \node(str)[right=of init,draw, dashed,align=center]{Sampled
    Follower's\\ strategy (Fig.~\ref{fig:method-outline}, Step~1)};

  \node[decision,below=of init,inner sep=-.4em] (attackerresponse) {Does better\\
    Follower's strategy\\ exist?};
  \node[process,right=of attackerresponse,
  mnemonic=Feasibility pass,dsmutate,xshift=-.95em,inner sep=2pt] (antibas) {Adjust
    Leader's strate-\\gy to lower
    $\pi_F^b$ payoff\\ compared to $\pi_F^r$ ($\ddagger$)};

  \draw[transition] (attackerresponse)--(antibas) node[pos=0.4,align=center,below]
  {Found\\ $\pi_F^b$};

  \draw[transition] (init) -- (attackerresponse) node (xx) [pos=0.5]{};
  \draw[transition] (antibas) -- ++(0,4em) -| (attackerresponse);

  \node[decision, below=of attackerresponse] (stopcon) {Stop
    condition?};

  \draw[transition] (attackerresponse)--(stopcon) node[pos=0.5,auto](snf)
  {Not found};
  \node[process,below=of stopcon,dsmutate,mnemonic=Positive pass] (positive) {Improve Leader's
    payoff\\against $\pi_F^r$ ($\dagger$)};

  \draw[transition] (stopcon) -- (positive) node[pos=0.5,auto] {No};
  \draw[transition] (positive) -- ++(-8em,0) |- (attackerresponse);

  \node[process,right=of positive, yshift=4em, xshift=-1em] (return) {Return best feasible\\ Leader's
    strategy};

  \draw[transition] (stopcon) -- ++(7em, 0) |- (return) node[pos=0.1,auto] {Yes};

  \node[process,right=of snf,xshift=-1.9em] (best) {Store best Leader's strategy};

  \draw[dataflow] (snf) -- (best);
  \draw[dataflow] (best) -- ++(0, -3em) -| (return);

  \draw[dataflow] (str) -- (init) node[pos=.5,above] {$\pi_F^r$};

  \node(as)[below=of
  return,draw,dashed,align=center,yshift=1em]{Fig.~\ref{fig:method-outline},
  Step~3};
  \draw[dataflow] (return) -- (as) node[pos=0.5] {Leader's payoff};
\end{tikzpicture}
  \caption{An overview of the method of finding Leader's mixed strategy corresponding to the requested Follower's strategy. Procedures marked in red adjust the current Leader's strategy. Blue labels refer to the names of procedures used in the text. Dashed boxes indicate connection points to the sampling procedure depicted in Fig.~\ref{fig:method-outline}.}%
  \label{fig:mixed-method-overview}
\end{figure}
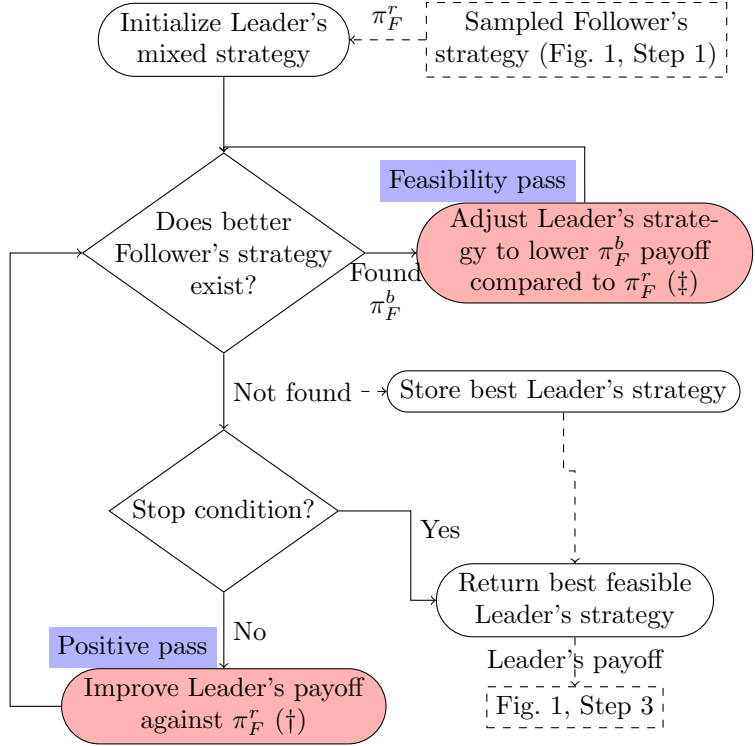

The Leader's strategy tree is initialized as a single path
representing a move sequence maximizing the Leader's payoff against
$\pi^r_F$. 
This sequence is found based on a limited number of UCT simulations.
All adjustments to the Leader's strategy are performed on a common (continuously evolving) tree-based representation. In Fig.~\ref{fig:mixed-method-overview} there are two procedures in which these strategy updates occur: \emph{feasibility pass} and \emph{positive pass}. The first one is executed when the current Leader's strategy becomes infeasible, i.e. there exists $\pi_F^b$ that yields higher Follower's payoff than $\pi_F^r$. The latter one is run to improve the Leader's payoff in the case of feasible (Leader's) strategy.
%
In both cases the same procedure, presented in Algorithms~\ref{alg:tree-walk} and~\ref{alg:momentum-adjustment}, is applied to update the Leader's strategy tree. The only difference lies in a move assessment subroutine which is explained below.

Algorithm~\ref{alg:tree-walk} starts off from the root
  node of the Leader's strategy tree and recursively descends to every
  leaf node of the tree. In each step of recursion one node of the tree, denoted by $n_c$, is processed. While in $n_c$,
  the algorithm first recursively calls itself for all child nodes of
  $n_c$. Once adjustment of these nodes (changes in probabilities of available moves) is completed, if there exist
  moves that are available to play from the IS corresponding to $n_c$ in
  the game, but not represented in the tree, a node representing one
  of them is added to the tree with some probability (equal to $0.3$ in the experiments) together with a path
  expanded from this newly-added node until a leaf node.
  Next, Algorithm~\ref{alg:momentum-adjustment} is applied to $n_c$.
%
\begin{algorithm}[t]
  \KwData{$n_c$ -- a node of a strategy tree currently processed,\phantom{aaa} $M$ -- a set of all moves that are added to the tree in $n_c$, $I_k$ -- an IS corresponding to $n_c$, $A_{I_k}$ -- a set of all moves in $I_k$.}
  \SetAlgoLined
  \ForEach{$m \gets M$}{
    recursively adjust successor of $n_c$ after playing $m$ against
    $\pi_F^b$ in \emph{feasibility pass} or $\pi_F^r$ in \emph{positive
      pass}
  }
  \If{$(rand()\le 0.3) \wedge \ (A_{I_k} \setminus M \not = \emptyset)$}{expand tree with one move $a'\in A_{I_k} \setminus M$}
  perform adjustment of $n_c$ as described in
  Algorithm~\ref{alg:momentum-adjustment}\;
  \caption{Strategy tree adjustment procedure}%
  \label{alg:tree-walk}
\end{algorithm}
\begin{algorithm}
  \SetAlgoLined

  \KwData{$prob\in {[0,1]}^M$ -- a vector of probabilities,
    $mom\in \mathbb{R}^M$ -- a momentum vector, $w\in \mathbb{R}$ -- a momentum
    normalization factor, $as\in \mathbb{R}^M$ -- an assessment vector. In each vector the $i$-th position corresponds to the $i$-th move.}

  $mom \gets mom + as$\;
  $w \gets w + L_1(as)$\;

  $prob \gets \max\{prob+mom/w, 0\}$\tcp{independent max at each
    position}

  $prob \gets normalizeOrEqual{prob}$\tcp{Normalize vector values so
    their sum is $1$ or, as a fallback, assign equal probability at each position in case all positions equal $0$}

  \caption{Node adjustment with momentum}%
  \label{alg:momentum-adjustment}
\end{algorithm}

The role of Algorithm~\ref{alg:momentum-adjustment}
is to accumulate direction of strategy changes in two passes: \emph{positive} and
\emph{feasibility}. A \emph{momentum} vector is used to store the resultant strategy adjustment stemming from those two passes. The algorithm uses a node \emph{assessment vector} ($as\in \mathbb{R}^M$) to indicate a direction of adjustment of the Leader's strategy and implements this adjustment based on the resultant direction accumulated in all previous iterations in a \emph{momentum vector} ($mom\in \mathbb{R}^M$). First, the momentum vector is updated by adding the assessment vector. A positive value in the assessment vector results in increasing the preference for the respective move, a negative one results in decreasing this preference. Next, a \emph{normalization factor} ($w\in \mathbb{R}$) is increased by adding $L1$ norm of assessment vector, to confine $mom/w$ to interval $[-1,1]$. Then the vector of move probabilities is updated with normalized $mom$ values and normalized to represent a proper probability distribution.

The last element of the method of finding the Leader's strategy is calculation of the assessment vector ($as$) used in Algorithm~\ref{alg:momentum-adjustment}, which is pass-dependent.
\begin{itemize}
\item In \emph{positive pass} the goal is to maximize the Leader's payoff. Consequently, $as$ value for move $a_i$: $as_i=\mathbb{E}U^L_{\pi_F^r}(a_i)-\mathbb{E}U^L_{\pi_F^r}$ is a difference between the Leader's expected payoff when move $a_i$ is played in the current state and an expected payoff when moves are played according to the current probabilities (Leader's mixed strategy). The higher the result of playing $a_i$ compared to the expected result arising from the current probabilities, the greater the $as_i$ value.
\item In \emph{feasibility pass} the goal is to modify the Leader's strategy in a way that $\pi_F^r$ will become the corresponding best response strategy. Hence, if the current IS is reachable when playing against both $\pi_F^r$ and $\pi_F^b$, then $as_i=(\mathbb{E}U^F_{\pi_F^r}(a_i)-\mathbb{E}U^F_{\pi_F^b}(a_i)) - (\mathbb{E}U^F_{\pi_F^r}-\mathbb{E}U^F_{\pi_F^b})$, i.e. $as_i$ is higher for moves that give better result against $\pi_F^r$ than against $\pi_F^b$.\\
    \noindent
If the current IS is reachable only when playing against $\pi_F^b$, then $as_i=\mathbb{E}U^F_{\pi_F^b} -
\mathbb{E}U^F_{\pi_F^b}(a_i)$. Note that an order of subtraction operands is reversed compared to a similar equation in the \emph{positive pass}. Here, the weaker the Follower's payoff when the Leader plays move $a_i$, the higher the value of $as_i$ because the goal is to discourage the Follower from playing strategy $\pi_F^b$.
\end{itemize}

\subsubsection{Stopping condition}%
\label{sec:stop-condition}
The algorithm depicted in Fig.~\ref{fig:mixed-method-overview} stops when
one of the following conditions is reached:
a number of executions of step ($\dagger$) exceeds $L_{max}=5000$, an improvement of the Leader's payoff in 500 subsequent iterations is less than $\varepsilon_I = 10^{-5}$, or a number of subsequent executions of step ($\ddagger$) without going to step ($\dagger$) exceeds $M_{max}=10000$ (\emph{infeasible strategy}).
%
%
%
Values of all steering parameters were selected based on a limited number of preliminary tests.

\subsection{Follower's strategy oracle}%
\label{sec:oracle}

Implementation of the above-mentioned algorithm requires an ability to find the Follower's strategy that yields better Follower's payoff against the current Leader's strategy.

The most straightforward approach, suitable for any EF game, would be to iterate through all possible Follower's strategies and choose the one with the highest expected payoff. Such an approach, however, is excessively slow and in practice hinders application of the method to games longer than $4$ steps. In order to address this issue our
implementation avoids enumeration of all Follower's strategies in the following way:
\begin{itemize}
\item Between any two consecutive questions to the oracle about the best response strategy, a collection of $Q$ pairs $(n, \pi_F)$ is maintained, where $n$ is the use counter and $\pi_F$ is the Follower's pure strategy ($Q=50$ was used in the experiments).
\item When asked about a better
Follower's strategy the algorithm first iterates over Follower's strategies from the above-mentioned collection. If there exists a strategy that yields better Follower's payoff than $\pi_F^r$ by more than $\varepsilon_O$ (equal to $10^{-2}$ in the experiments), then this strategy is returned and its use counter is incremented. Otherwise:
  \begin{enumerate}
  \item If the collection is filled up (contains $Q$ strategies), a strategy with the smallest use counter is removed.
  \item Full enumeration of the Follower's strategies is performed and the best one is selected and added to the collection with $n=1$.
  \end{enumerate}
\end{itemize}
%
%
%
\section{Experimental evaluation}%
\label{sec:experiments}

Evaluation of \emph{O2UCT} was performed on three game sets: Warehouse Games proposed in~\cite{KarwowskiMandziuk_2019}, its modified version with more diverse payoffs, and Search Games used in~\cite{bosansky2015}.
\subsection{Warehouse Games (WHG)}
\label{sec:warehouse-games}
WHG
model interactions between the attacker
and the defender
in a warehouse/office building. A game graph includes the following three types of distinguished vertices: one defender's starting point, one attacker's starting point, and several asset locations (targets).
%
%
The game is sequential and in each turn each of the players can either stay in the current vertex or move to any adjacent vertex. A full description of a game model is presented in~\cite{KarwowskiMandziuk_2019}. In the experiments $25$ WHG instances were used.

\subsection{Modified Warehouse Games (WNZ)}
\label{sec:warehouse-games-NZ}
WHG setting is relatively close to zero-sum games, the average Pearson's correlation between the Leader's and the Follower' payoffs equals $-0.82$. To provide a more challenging setting we used a game generator from~\cite{KarwowskiMandziuk_2019} to obtain benchmark games with more diverse payoffs.
The following ranges for uniform distributions were applied: attacker's penalty in targets: $[-1,0.2]$ and regular vertices: $[-1,0]$ (when caught by the defender), attacker's reward in targets: $[-0.2,1]$ and the corresponding defender's penalty in targets: $[-1,0.2]$ (successful attack). Defender's reward (for catching the attacker) in non-target and target vertices was fixed at $0.1$ and $0.2$, resp. If the game ended due to reaching the round limit, with no interception of the attacker or reaching a target by him/her, a neutral payoff of $0$ was assigned to both players. A set of $25$ games was generated with the above parameters consisting of games that are less zero-sum like -- the average Pearson's correlation for this set equals $-0.57$. An example WNZ graph is depicted in Fig.~\ref{fig:game-graph-gr}.

\begin{figure}
  \centering
    \includegraphics[width=0.80\columnwidth]{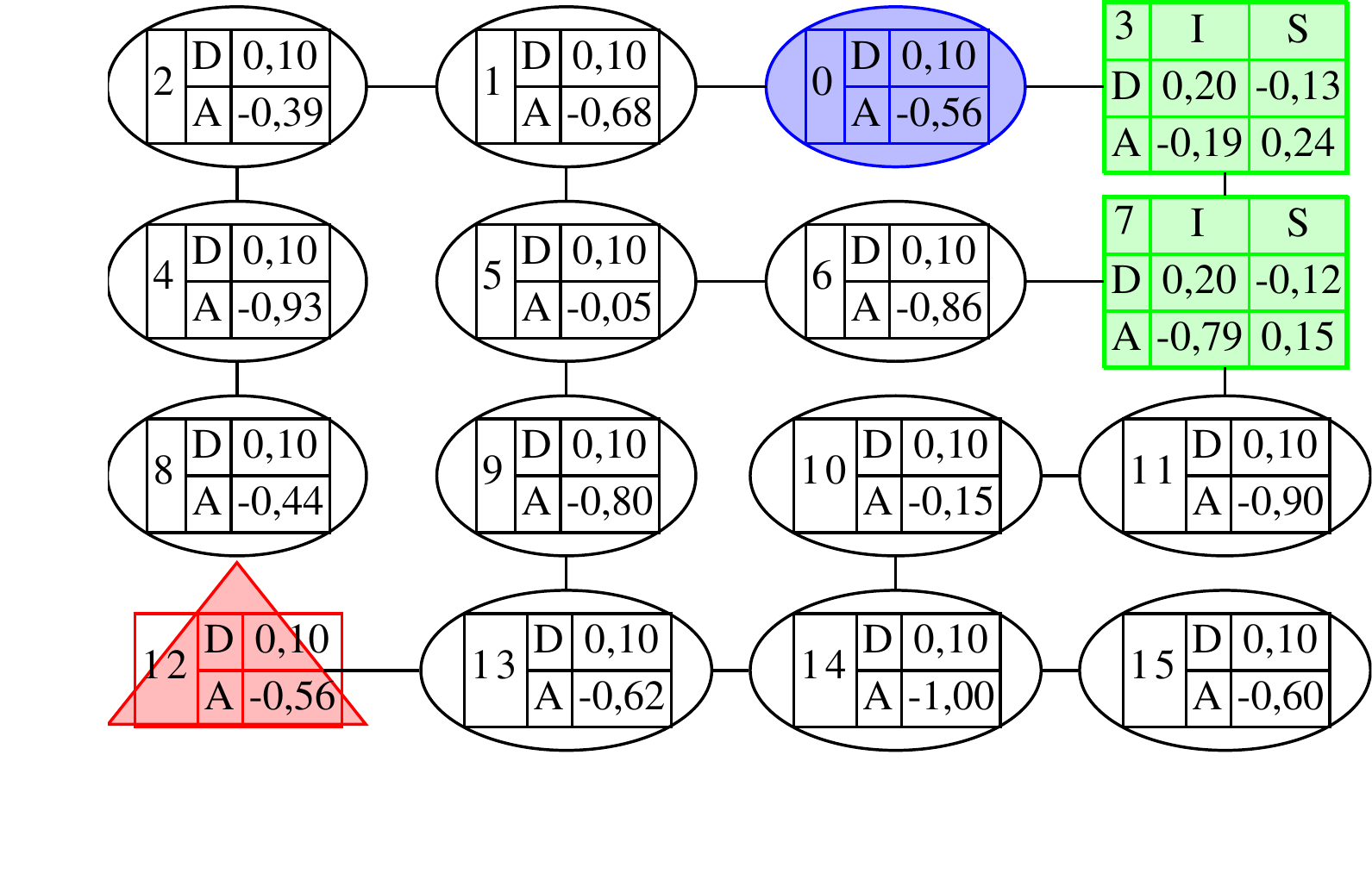}
    \caption{Example WNZ graph. All games are defined on a $4\times 4$ grid.
    Rectangular vertices are targets, a triangle vertex is evader's starting point, a blue circle vertex is defender's starting point. Values denote payoffs for the evader and the defender, resp.\ in the case of evader's interception in a given vertex. Additional utilities, in case of successful attack, are assigned in targets (the second column).}
    \label{fig:game-graph-gr}
\end{figure}
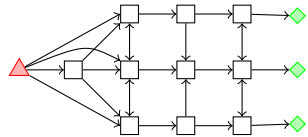
\begin{figure}
    \centering
  \begin{tikzpicture}[
    regular/.style={draw},
    entry/.style={draw, regular polygon, regular polygon sides=3,
      inner sep=0.15em,red, fill=red!30},
    target/.style={draw,diamond,green, inner sep=0.15em, fill=green!30},
    node distance=1.4em,
    winning/.style={solid},
    every node/.style={solid}
    ]

    \node(e) [entry] {};
    \node(x) [regular, right=of e] {};
    \node(a2) [regular, right=of x] {};
    \node(b2) [regular, right=of a2] {};
    \node(c2) [regular, right=of b2] {};
    \node(d2) [target, right=of c2] {};

    \node(a1) [regular, above=of a2] {};
    \node(b1) [regular, above=of b2] {};
    \node(c1) [regular, above=of c2] {};
    \node(d1) [target, above=of d2] {};

    \node(a3) [regular, below=of a2] {};
    \node(b3) [regular, below=of b2] {};
    \node(c3) [regular, below=of c2] {};
    \node(d3) [target, below=of d2] {};

    \draw[->] (e) -- (x);
    \draw[->,winning] (e) -- (a1.west);
    \draw[->,winning] (e) .. controls ++(2.5em, 1em) ..  (a2.north west);
    \draw[->,winning] (e) -- (a3.west);

    \draw[->] (x) -- (a1);
    \draw[->] (x) -- (a2);
    \draw[->] (x) -- (a3);

    \draw[->,winning] (a1) -- (b1);
    \draw[->,winning] (a2) -- (b2);
    \draw[->,winning] (a3) -- (b3);

    \draw[->,winning] (b1) -- (c1);
    \draw[->,winning] (b2) -- (c2);
    \draw[->,winning] (b3) -- (c3);

    \draw[->,winning] (c1) -- (d1);
    \draw[->,winning] (c2) -- (d2);
    \draw[->,winning] (c3) -- (d3);

    \draw[<->] (a1) -- (a2);
    \draw[->] (b1) -- (b2);
    \draw[<->] (c1) -- (c2);

    \draw[<->] (a3) -- (a2);
    \draw[->] (b3) -- (b2);
    \draw[<->] (c3) -- (c2);

  \end{tikzpicture}
  \caption{Search Games graph.}
  \label{fig:Search}
 \end{figure}

\subsection{Search Games (SEG)}
\label{sec:search-games}

SEG instances were built on a graph presented in Fig.~\ref{fig:Search} previously used in~\cite{bosansky2015}. Five sets of payoff values in the target vertices for each of the two  variants of an attacker's mobility restrictions (in the first one the attacker can wait in a vertex, in the other one he/she is forced to move in each round) were generated. In each case, game variants with $T=4,5$ and $6$ were considered leading to $30$ test instances in total.

Similarly to~\cite{bosansky2015}, the following distributions of payoffs were used in the tests. In case of catching the attacker  the reward for the defender equaled $1$ and the penalty for the attacker was equal to $-1$. The attacker's rewards in targets in case of a successful attack were sampled uniformly from $[1, 2]$ and the respective defender's penalty was equal to $-1$. Otherwise, when the game ended due to reaching the step limit $T$, a neutral payoff equal to $0$ was assigned to each side.
%

\subsection{Experimental setup}%
\label{sec:games}

In the following description \emph{BC2015} refers to the method from~\cite{bosansky2015}, \emph{C2016} to approach from~\cite{cermak2016using} (a variant AI-MILP was used) and \emph{CBK2018} to the method from~\cite{CBK2018} which, following suggestions from the authors, is implemented in two variants: (a): $\epsilon=0.3, \delta=0.4$ and (b): $\epsilon=0.0, \delta=0.4$ (the latter provides better SSE approximations, though requires longer computation time).
The following experiments were performed to evaluate efficiency and scalability of \emph{O2UCT}.
\begin{itemize}
\item For each of WHG and WNZ instances (defined by game layout and $T$) $15$ trails of \emph{O2UCT} were run and for each SEG $5$ \emph{O2UCT} tests were run. Multiple trials were required due to stochastic nature of the method.
\item For each WHG, WNZ and SEG instance and each of the MILP-based methods (\emph{BC2015}, \emph{C2016}, \emph{CBK2018(a)}, \emph{CBK2018(b)}) a single trial was made (all methods are deterministic). Obtained results were used as a baseline for \emph{O2UCT} assessment.
\end{itemize}
All experiments were run on Intel Xeon Silver 4116 @ 2.10GHz with 256GB\ RAM. Experiments involving \emph{O2UCT} were run in parallel, each with 8GB\ RAM assigned. The remaining tests were run in sequential manner with full memory available to a single process. Each test was run with a time limit of $200$ hours and was forcibly terminated if did not finish within the allotted time or exceeded available memory.
\section{Results}
\label{sec:quality}

Performance of \emph{O2UCT} is analyzed in two dimensions: an expected Leader's payoff and time scalability. In both cases the results are presented separately for WHG, WNZ and SEG and grouped by the number of nodes of an extensive-form game $|\mathcal{S}\cup\mathcal{Z}|$:
  \begin{equation}
    \label{eq:buckets}
    bucket=10^{round(\log_{10}|\mathcal{S}\cup\mathcal{Z}|)},
  \end{equation}
where $round$ rounds a number to the nearest integer. Consequently, games are grouped by the orders of magnitude of game nodes. Such a grouping combines two sources of game complexity: the structure of an underlying game graph and the game length.
  \begin{figure*}
    \centering
    \includegraphics[width=.45\linewidth]{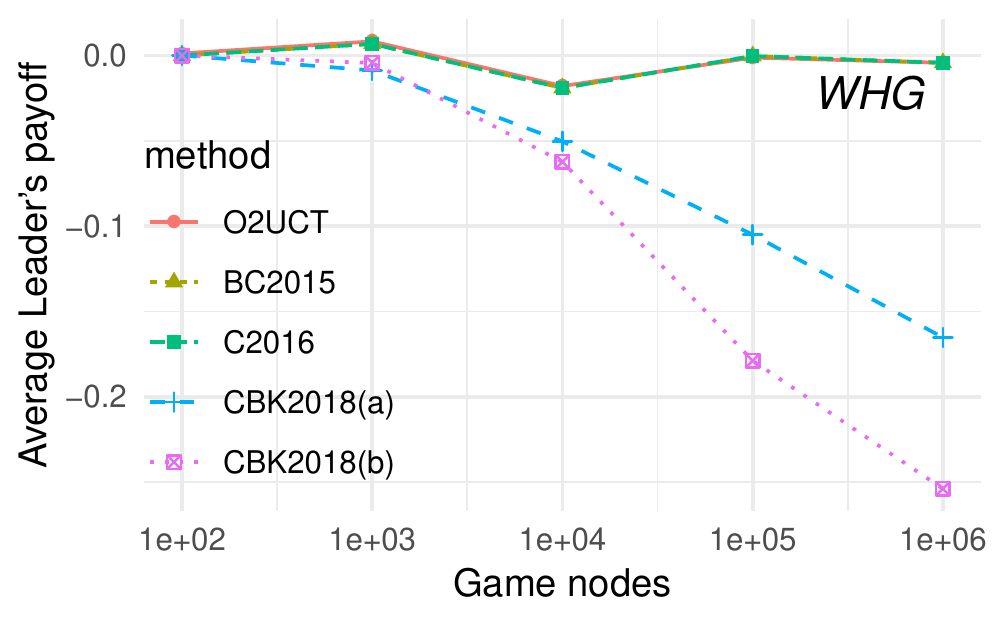}
    \qquad
    \includegraphics[width=.45\linewidth]{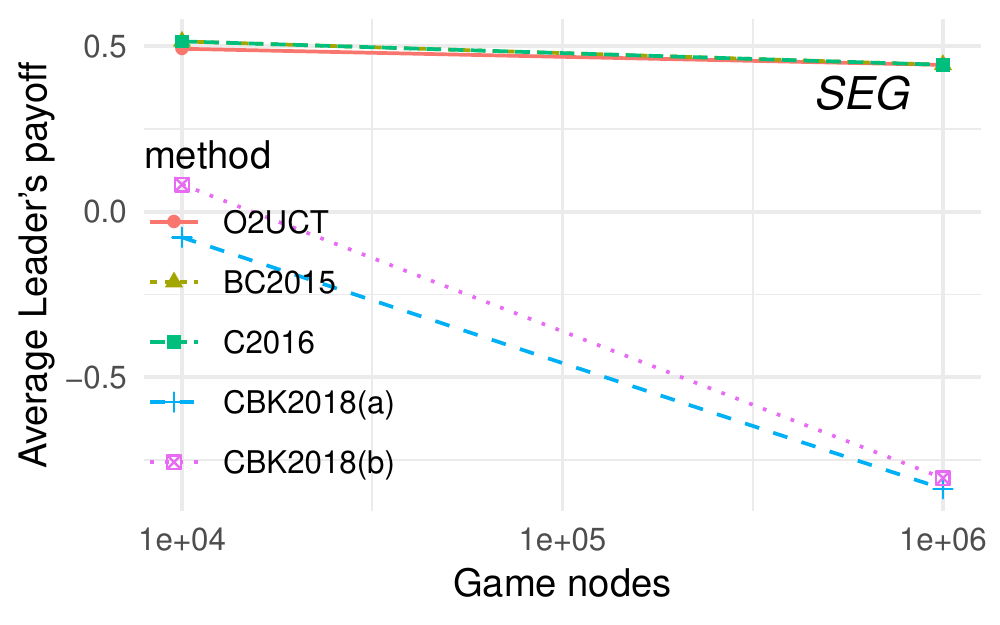}
    \qquad
    \includegraphics[width=.45\linewidth]{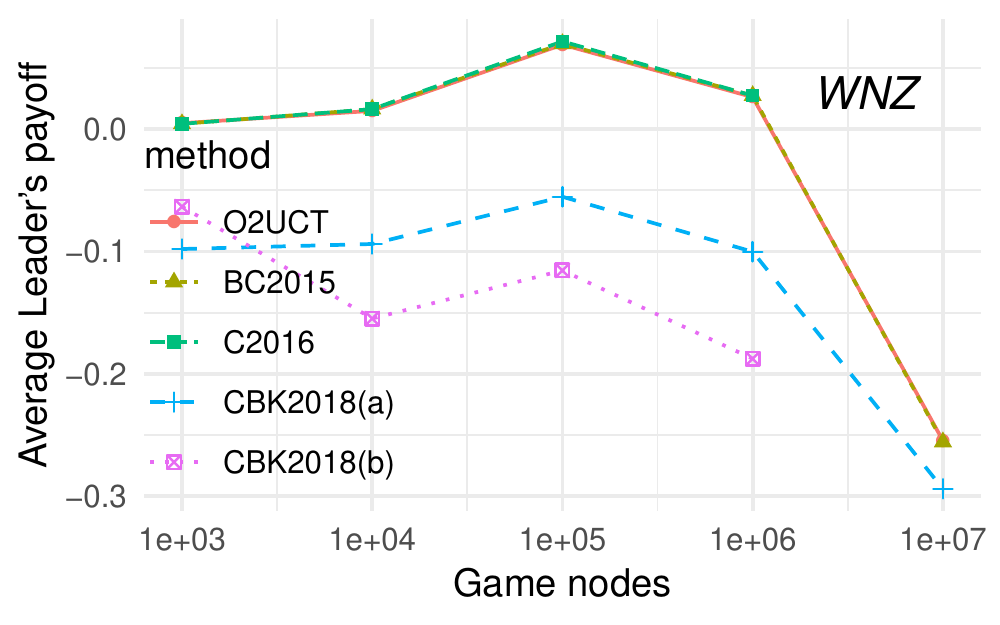}
    \caption{The average Leader's payoffs obtained by the tested methods for three benchmark sets  w.r.t. the number of game nodes. Plots are cut at the buckets for which exact methods were still able to solve at least $70\%$ of the respective game instances within $200h$ time limit. In these borderline cases the average payoffs are calculated for the subsets of solved games only.}
    \label{fig:quality}
  \end{figure*}

    \begin{figure*}[t]
    \centering
    \includegraphics[width=.45\linewidth]{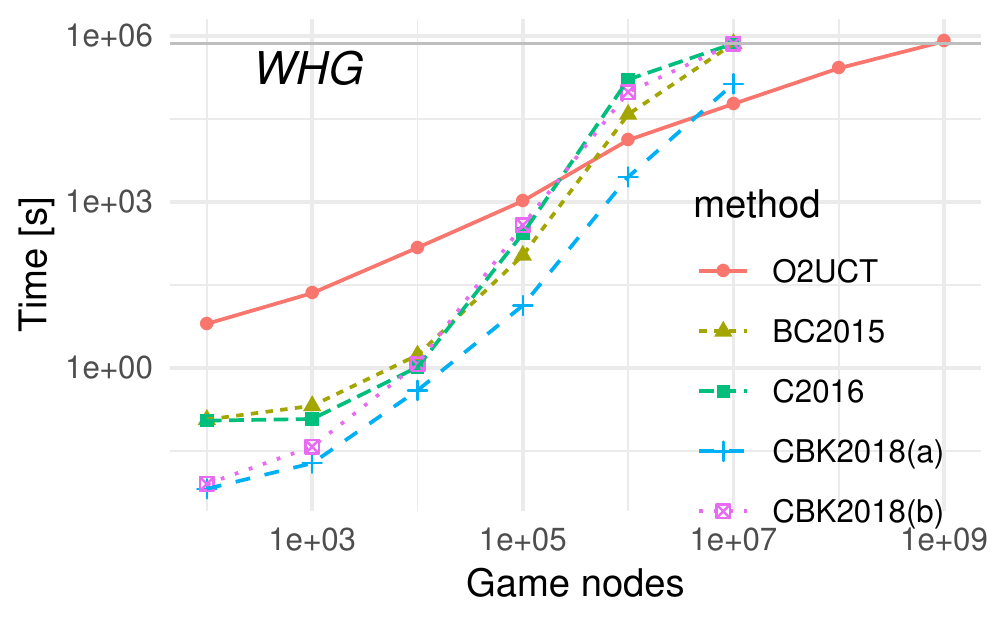}
    \qquad
    \includegraphics[width=.45\linewidth]{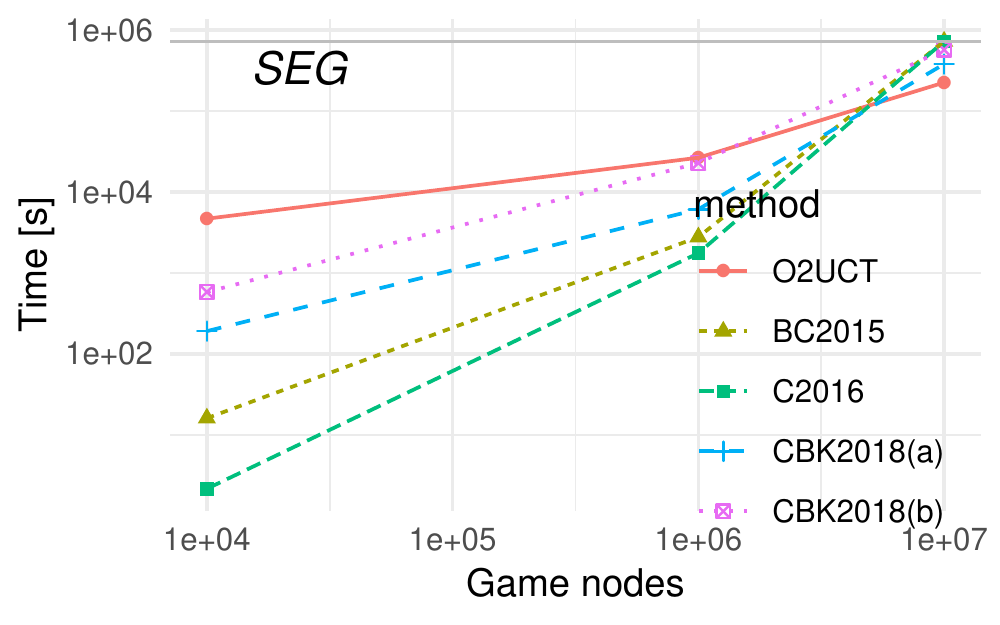}
    \quad
    \includegraphics[width=.45\linewidth]{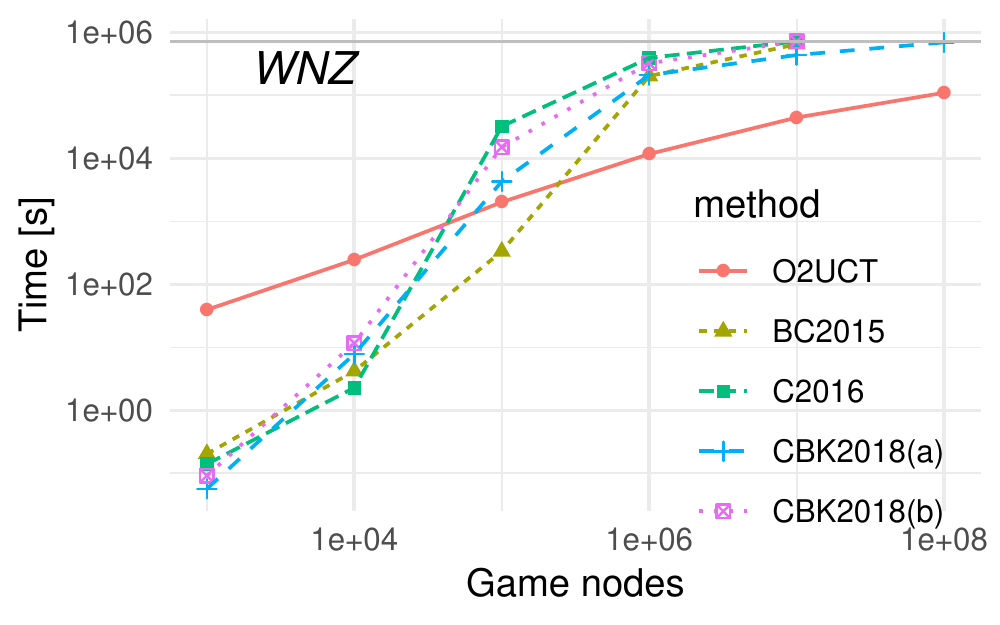}
    \caption{The average computation times for three benchmark sets. \emph{BC2015} and \emph{C2016} were unable to solve some WHG, SEG, WNZ game instances from $10^7, 10^7, 10^6$ buckets, respectively due to hitting the time limit of $200h$ (marked as a gray line). In such cases time limit value was used in place of the respective computation time.}
    \label{fig:graph-times}
  \end{figure*}

\subsection{Payoffs}%
\label{sec:payoffs}
Fig.~\ref{fig:quality} presents the Leader's payoffs averaged for all game instances in the respective benchmark sets, calculated at the end of each test against the worst-case Follower (found by enumerative check among all possible Follower's pure strategies). Particular points are plotted if at least $70\%$ of games from a given bucket were solved by a given method within allotted time. Otherwise the respective points are omitted as their comparison would be meaningless.
Plots of \emph{BC2015} and \emph{C2016} overlap as both refer to exact methods.
Generally speaking, the average Leader's payoffs calculated by \emph{O2UCT} are very close to optimal results while both variants of \emph{CBK2018} visibly diverge from optimal outcomes, specifically for larger games.

In summary, we believe that the quality of strategies (the Leaders's payoffs) found by \emph{O2UCT} are very encouraging, as the average results are only slightly worse than the optimal ones, even for the most complex games.

\subsection{Computation times}%
\label{sec:time}

Computation time analysis is presented in Fig.~\ref{fig:graph-times}. In the case of reaching computation time limit for a given game instance, a value of $200h$ (time limit) was used in place of the respective result (and averaged with the remaining times in the bucket).

For small WHG and WNZ instances \emph{O2UCT} is the slowest method, however
for larger games
it starts to outperform MILP approaches. For the $10^7$ bucket \emph{O2UCT} managed to solve all game instances (both WHG and WNZ) within allotted time, \emph{CBK2018(a)} solved $23$ and $13$ games of type WHG and WNZ, respectively, while the remaining methods reached the time limit in almost all cases.

For SEG games \emph{O2UCT} was the slowest for instances with up to $10^6$ nodes but for $10^7$ nodes its computation times are already the shortest since both \emph{CBK2018} variants hit the time limit in at least $40\%$ of the cases and \emph{BC2015} and \emph{C2016} did not solve a single instance.
%
%
%
%
%
%

In summary, it can be observed in Fig.~\ref{fig:graph-times} that for small games the results are in favor of MILP approaches, but for larger games MILP methods scale poorer than \emph{O2UCT}.
%
While all considered methods scale exponentially, extrapolation of results to yet bigger games suggests that \emph{O2UCT} is the best-scaling methods amongst the tested ones.


At the same time, it shouldn't be forgotten that \emph{BC2015} and \emph{C2016} are exact methods that yield theoretically guaranteed SSE utilities, while \emph{O2UCT} is only experimentally proven to yield optimal or close-to-optimal strategies.
%

While exact measurements of memory usage were not performed (it was
not possible because of using Java Virtual Machine and its garbage
collection facilities) we noted that \emph{O2UCT} was able to compute results
for $10^9$ game nodes using $8$GB of memory while
solver based methods started running out of ($256$GB)
memory for games with $10^7$ nodes.

\section{Conclusions}%
\label{sec:conclusions}

This paper presents a novel double-oracle approach for approximating SSE strategy in sequential games with imperfect information and perfect recall. The method does not rely on solving LP/MILP (which is the most common approach) but consists in iterative MCTS/UCT sampling of the Follower's strategy space alternated with adequate modification of the Leader's behavior strategy.

Experimental evaluation shows that proposed approach provides high-quality solutions (optimal in vast majority of the tests) and scales visibly better than state-of-the-art MILP-based methods used for reference. Moreover \emph{O2UCT} requires substantially less memory resources and is therefore capable of solving more complex game instances. Lower memory requirement stems from two factors: application of a double oracle approach which does not require storing in memory all possible strategy profiles simultaneously and dynamic expansion of the Leader's strategy tree -- an approach similar in some aspects to the idea of column generation in LP methods. However, what makes \emph{O2UCT} distinct from column generation is the use of a game-independent UCT metaheuristic (instead of a game-specific heuristic) when searching for the most promising moves.

Good time and memory scalability of \emph{O2UCT} enables its application to larger (than in the case of other approaches) game instances. Furthermore, iterative nature of the method allows an easy adjustment of a balance between computation time and quality of results. The outer sampling procedure employs UCT, which is an anytime algorithm that can be stopped in any moment, though still returning a high quality solution (the best one found so far). An anytime property makes \emph{O2UCT} particularly well suited to problems with strictly allotted time for finding the Leader's strategy.

The introduced method can be applied to any sequential game as it does not depend on any specific game structure or game property.
\bibliographystyle{apalike}  
\bibliography{mixedse}

\end{document}